\documentclass[prl,preprint,superscriptaddress,showpacs,byrevtex]
{revtex4}
\date{\today} 
\draft 
\tighten 
\newcommand{\be}{\begin{equation}} 
\newcommand{\ee}{\end{equation}} 
\newcommand{\bea}{\begin{eqnarray}} 
\newcommand{\eea}{\end{eqnarray}} 
\usepackage[dvips]{graphicx} 

\catcode`@=11 \@addtoreset{equation}{section} \catcode`@=12 

\begin{document} 
\def\sqr#1#2{{\vcenter{\hrule height.3pt 
ÊÊÊÊÊ \hbox{\vrule width.3pt height#2ptÊ \kern#1pt 
ÊÊÊÊÊÊÊÊ \vrule width.3pt}Ê \hrule height.3pt}}} 
\def\square{\mathchoice{\sqr67\,}{\sqr67\,}\sqr{3}{3.5}\sqr{3}{3.5}} 
\def\today{\ifcase\month\or 
Ê January\or February\or March\or April\or May\or June\or July\or 
Ê August\or September\or October\or November\or December\fi 
Ê \space\number\day, \number\year} 

\def\Bbb{\bf} 
\topmargin=-0.3in


\newcommand{\ww}{\mbox{\tiny $\wedge$}} 
\newcommand{\pp}{\partial} 

\title{$J/\Psi$ Production in pp Collisions at $\sqrt s$ = 200 GeV at RHIC} 

\author{Fred Cooper} \email{fcooper@nsf.gov}
\affiliation{ National Science Foundation, Arlington, VA 22230, 
 USA and T-8, Theoretical Division, Los Alamos National Laboratory,
Los Alamos, NM 87545, USA} 
\author{Ming X. Liu} \email{mliu@lanl.gov}
\affiliation{
P-25, Physics Division, Los Alamos National Laboratory,
Los Alamos, NM 87545, USA} 
\author{Gouranga C. Nayak} \email{nayak@insti.physics.sunysb.edu} 
\affiliation{
C. N. Yang Institute for Theoretical Physics, Stony Brook University, 
SUNY, Stony Brook, NY 11794-3840, USA }

\begin{abstract} 
We study $J/\psi$ production in pp collisions at RHIC  within the PHENIX detector acceptance range  using  the color singlet 
and color octet mechanism which are based on  pQCD and NRQCD. Here we show that the color 
octet mechanism reproduces the RHIC data for $J/\psi$ 
production  in pp collisions  with respect to the  $p_T$ distribution, the
rapidity distribution and the  total cross section 
at $\sqrt s$ = 200 GeV. The color singlet mechanism leads to a relatively small contribution to the total cross section when compared to the 
octet contribution.
\end{abstract} 
\pacs{ }
\maketitle 

Understanding the  $J/\psi$  production mechanism in high energy hadronic 
collisions is an important topic in QCD and in collider physics. Two prominent  mechanisms for 
$J/\psi$  production at high energy colliders are 1)the  color singlet mechanism \cite{sing,org}
and 2)the color octet mechanism \cite{oct}, both of which rely on pQCD and NRQCD. 
At Tevatron energies, the color singlet mechanism was found to give too small a yield and the 
color octet mechanism for $J/\psi$  production was then introduced \cite{braaten,cho2}. It was 
shown that the color octet mechanism, with both parton fusion and parton fragmentation 
processes included, reproduced the Tevatron data \cite{cdf}. 
The relativisitic heavy ion collider (RHIC) 
at Brookhaven is a unique facility which can collide unpolarized and polarized protons at 
$\sqrt s$ =200 and 500 GeV as well as two heavy nuclei such as gold on gold. 
It has also the 
capability of exploring light-heavy interactions such as the collisions pAu and dAu 
at $\sqrt s$ = 200 GeV. 
The unpolarized pp collision data  serve as a 
baseline 
for understanding collective phenomena such 
as the production of 
a quark 
gluon plasma that is expected in the heavy ion collisions as well 
as 
providing the necessary information needed to extract interference 
terms generated when 
considering polarized pp collisions. One of the 
key things to be measured in Au Au collisions 
as well as in polarized 
pp collision is $J/\psi$ production. In relativistic  heavy ion collisions,
$J/\psi$  suppression has 
been posited to be one of the major signatures of the production of 
the quark-gluon plasma \cite{nayak}. 
In case of polarized pp collisions the measurement 
of $J/\psi$ production 
is an important ingredient in extracting the 
polarized gluon distribution function for the proton 
directly. For that 
reason, understanding the dominant mechanism of $J/\psi$ production
is important. Thus it is crucial to first determine the 
relative importance of color singlet and 
color octet contributions in 
the existing unpolarized proton-proton scattering data at RHIC energies.
In this paper we will study  $J/\psi$  production in pp collisions at RHIC 
by using the color 
singlet and color octet mechanism within pQCD and NRQCD. We 
will study the $p_T$ distribution, 
rapidity distribution and total cross 
section of the  $J/\psi$  production at $\sqrt s$ = 200 GeV 
pp collisions 
Êand compare those with the recent measurements by the PHENIX 
collaboration at 
RHIC \cite{phenix}. 

In the color singlet mechanism,  quarkonium is 
formed as a non relativistic 
bound state of  a heavy quark-antiquark pair via static gluon exchange. 
In this mechanism it is assumed that the   $ Q\bar Q$ pair is produced 
in the color singlet state at the production point with appropriate 
spin (S) and orbital angular momentum 
quantum number (L) and then evolves into a bound state 
$^{2S+1}L_J$ 
with total angular momentum quantum number (J). 
The relative momentum of the $Q\bar Q$ pair inside 
the quarkonium is assumed to be small compared to the mass $m$ of 
the heavy quark so that the $Q$ and $\bar Q$ will not fly apart 
to form heavy mesons. The non-relativistic non-perturbative wave functions 
and its derivatives appearing in the color singlet calculation 
are either determined from the potential model or 
taken from experiment. 

The leading term in the $p_T$ distribution  for the  cross section for  $J/\psi$  production first occurs at order $\alpha_s^3$. 
The $p_T$ distribution for heavy quarkonium  production 
in the color singlet mechanism is given by: 
\bea 
&& \frac{d\sigma}{dp_T }(AB\rightarrow J/\psi, \chi_J + X)~ 
=~\sum_{a,b} \int dy \int dx_a ~x_a \nonumber \\ 
&& f_{a/A}(x_a, Q^2)~ 
x_b^{\prime} f_{b/B}(x_b^{\prime}, Q^2)~ 
\frac{2 p_T}{{x_i -\frac{M_T}{\sqrt{s}} e^{y}}} \nonumber \\
&& \frac{d \hat{\sigma}}{{d\hat{t}}}  
\left(ab 
\rightarrow( ^{2S+1}L_J )c\right), 
\label{dsdp1} 
\eea 
where 
\be 
x_b^{\prime}=\frac{1}{\sqrt{s}} \frac{x_b \sqrt{s}M_T 
e^{-y}-M^2}{x_b \sqrt{s}-M_T e^{y}}. 
\label{x2f} 
\ee 
In the above equation $a,b,c$ are light quarks and gluons. 
$f(x,Q^2)$ is the parton distribution function at longitudinal 
momentum fraction $x$ and at factorization scale $Q$.Ê 
$M$ is the mass of the bound state quarkonium 
and 
$M_T=\sqrt{p_T^2+M^2}$. The partonic level differential 
cross section 
$\frac{d \hat{\sigma}}{{d\hat{t}}}(ab \rightarrow (^{2S+1}L_J)c)$ 
contain the non-relativistic 
wave function 
$|R(0)|^2$ (for direct $J/\psi$ process) 
and its derivatives $|R^\prime(0)|^2$ (for $\chi_J$ processes) at the origin \cite{sing}. For the non relativistic wave functions at the origin 
we take 
the Buchmuller-Tye wave function  \cite{bm}. 
The numerical value is \cite{bm}: $|R(0)|^2$=0.81 GeV$^3$. 
For the derivative of the radial wave function at origin we use 
\cite{brat}: $\frac{9}{2\pi}\frac{|R^\prime(0)|^2}{M_c^4}$=15 MeV. 

In the color octet mechanism the relativistic 
effects are taken into account which are neglected in the color singlet case. 
In the color octet mechanism using 
an effective 
field theory called non-relativistic QCD (NRQCD), the dynamical 
gluon enters 
into 
Fock state decompositions of the quarkonium states. 
In NRQCD the expansion is carried out in terms of the relative velocity 
$v$ ($v^2 \sim$ 0.23 for $ C \bar C$ system and 0.1 for $ B \bar B$ system) 
of the $Q \bar Q$ bound state. The NRQCD Lagrangian density is given by 

\be 
{\cal L}_{NRQCD}= {\cal L}_{light} + {\cal L}_{heavy} + {\cal L}_{correction}. 
\ee
In the above equation light refers to light quarks and gluons, heavy refers to 
lowest order non-relativistic heavy quarks part and correction is higher order corrections
in heavy quark sector.
The explicit expressions for the three terms can be found in \cite{braaten}.
In NRQCD the dynamical gluons enter into the Fock state decompositions 
of different physical states. The wave function of an 
S-wave orthoquarkonia state 
$|\psi_Q>$ is expanded as follows: 
\bea 
&&{|\psi_Q>} = O(1) |Q\bar Q[^3S^{(1)}_1]>+ 
O(v) |Q\bar Q[^3P^{(8)}_J]g> \nonumber \\ 
&&+ O(v^2) |Q\bar Q[^3S^{(1,8)}_1]gg>+ 
O(v^2) |Q\bar Q[^1S^{(8)}_0]g> \nonumber \\ 
&&+ O(v^2) |Q\bar Q[^3D^{(1,8)}_J]gg>+ ..... 
\label{jp} 
\eea 
and the wave functions of a P-wave orthoquarkonium state $|\chi_{QJ}>$ has a similar expansion: 
\bea 
|\chi_{QJ}> = O(1) |Q\bar Q[^3P^{(1)}_J]>+ O(v) |Q\bar Q[^3S^{(8)}_1]g> 
\nonumber \\ 
+O(v^2) .... 
\label{cj} 
\eea 
In the above equation (1,8) refers to singlet and octet state of the 
$Q\bar Q$ pair. 
After a $ Q \bar Q$ is formed in its color octet state 
it may absorb a soft gluon and transform into 
$|\chi_{QJ}>$ via eq. (\ref{cj}) and then becomeÊ a  
$J/\psi$  by photon decay. 
The $Q \bar Q$Ê pairÊ in the color octet state can also 
emit two long wavelength gluons 
and become a  $J/\psi$  via eq. (\ref{jp}) and so on. 
The non-perturbative 
matrix elements 
can be fitted from other experiments or can be determined 
from  lattice calculations. 

In the color octet mechanism the $p_T$ distribution  differential cross section for  
$J/\psi$  production is given by: 
\bea 
&&\frac{d\sigma}{dp_T }(AB\rightarrow \psi_Q (\chi_J) +X)~ 
=~\sum_{a,b} \int dy \int dx_a ~x_a \nonumber \\ 
&&f_{a/A}(x_a, Q^2)~ x_b^{\prime} f_{b/B}(x_b^{\prime}, Q^2)~ 
Ê\times \frac{2 p_T}{{x_a -\frac{M_T}{\sqrt{s}} e^{y}}} \nonumber \\ 
&& \frac{d \hat{\sigma}}{{d\hat{t}}}(ab \rightarrow C\bar C[^{2S+1}L_J^{(8)}]c 
\rightarrow \psi_Q (\chi_J)) 
\label{dsdp2}, 
\eea 
where the partonic level differential cross section is given by 
\bea 
&&\frac{d \hat{\sigma}}{{d\hat{t}}}(ab \rightarrow C\bar C[^{2S+1}L_J^{(8)}]c 
\rightarrow \psi_Q (\chi_J))Ê ~=\frac{1}{16 \pi {\hat s}^2} \nonumber \\ 
&& \Sigma ~|{\cal{A}} 
(ab \rightarrow C\bar C[^{2S+1}L_J^{(8)}]c)_{{\rm short}}|^2 
<0|{\cal{O}}_8^{\psi_Q (\chi_J)}(^{2S+1}L_J)|0> 
\eea 
In our calculation all the partonic level matrix elements squared 
$$ \Sigma ~|{\cal{A}} 
(ab \rightarrow C\bar C[^{2S+1}L_J^{(8)}]c)_{{\rm short}}|^2$$ 
and the non-perturbative matrix elements 
$$<0|{\cal{O}}_8^{\psi_Q (\chi_J)}(^{2S+1}L_J)|0>$$ 
are taken from 
\cite{cho2}. 

For the rapidity distribution and total cross section of  $J/\psi$  
production 
the leading  contribution is 
proportional to $\alpha_s^2$  and we will only consider that contribution here. 
In leading order,  the parton fusion processes contribute to the total 
cross 
section for $J/\psi$ ($\chi_J$) 
production in pp 
collisions as follows: 
\bea 
&& \sigma^{pp \rightarrow J/\psi (\chi_J)} = \Sigma_{a,b} \int dx_a \int dx_b 
f_{a/A}(x_a,Q^2) f_{b/B}(x_b,Q^2) \nonumber \\ 
&& \sigma^{ab \rightarrow J/\psi (\chi_J)}(\hat s) \delta (x_a x_b -4m^2/s) 
\eea 
where $\sigma^{ab \rightarrow J/\psi (\chi_J}(\hat s)$ is the partonic 
level cross section at leading order which are derived in \cite{fred1,cho2}. 
In the above equation 
$m$ is the mass of the charm quark. 
At the leading order not all the processes contribute to the 
total cross section in the 
singlet and octet channel. In the color singlet channel the  
$J/\psi$ production cross section at $\alpha_s^2$ order is given by: 
\begin{eqnarray} 
ÊÊ \sigma_1^{pp \rightarrow J/\psi}(s)= 
ÊÊÊÊ \sigma_1^{pp \rightarrow \chi_0}(s) 
BR_{\chi_0}, 
ÊÊÊÊ +\sigma_1^{pp \rightarrow \chi_2}(s) 
BR_{\chi_2}. 
\end{eqnarray} 

Here $BR_{\chi}$  refers to the branching ratio of a $\chi$ to decay into  a $J/ \psi$.
Similarly in the octet channel at the leading order 
the $J/\psi$ production cross section is given by: 
\begin{eqnarray} 
&&ÊÊ \sigma_8^{pp \rightarrow J/\psi}(s)= 
\sigma_8^{pp \rightarrow J/\psi}(s) 
ÊÊÊÊ +\sigma_8^{pp \rightarrow \chi_1}(s)Ê BR_{\chi_1}Ê \nonumber \\ 
&& +\sigma_8^{pp \rightarrow \chi_0}(s) BR_{\chi_0} 
+\sigma_8^{pp \rightarrow \chi_2}(s) 
BR_{\chi_2}. 
\end{eqnarray} 
In the above equations the direct $J/\psi$ production and the decay of 
$\chi_c$ with their 
correcsponding branching ratios to $J/\psi$ 
are included. 
In the leading order we use the same values of the non-perturbative 
matrix elements used in the singlet and octet channels as 
in \cite{losing}. 

We willÊ use the  GRV98\cite{grv98} and MRST99 
\cite{mrst99} parton distribution functions inside the 
proton. The factorization 
scales are set to be 
$Q=2m$ for the total cross section 
and rapidity distribution plots,  and $Q=\sqrt{p_T^2+M^2}$ for 
the plot of the  $p_T$ distribution. 

In Fig.1 we present the $p_T$ distribution of the  
$J/\psi$  production cross section separately for the color singlet and color octet mechanism 
including the effects
of cuts due to the PHENIX detector acceptance range. We have considered 
all the parton fusion processes \cite{cho2,cooper} in our calculation. 
In these plots we have used the GRV98 and MRST99 
parton distribution functions. As the non-perturbative NRQCD matrix elements has
uncertainties as given in \cite{cho2} we present two plots for each of the  PDF's which
correspond to the upper and lower limit on the non-perturbative matrix elements.
The solid line corresponds to the results obtained by using the GRV98 PDF's with lowest values of the
non-perturbative matrix elements as given in \cite{cho2}. 
The upper dashed line corresponds to the results obtained by using the GRV98 PDF's with highest 
values of the non-perturbative matrix elements. 
The upper dot-dashed line corresponds to the results obtained by using the MRST99 PDF's with highest values of the
non-perturbative matrix elements. 
The lower dot-dashed line corresponds to the results obtained by using the MRST99 PDF's with lowest values of the
non-perturbative matrix elements. 
The lower dashed line corresponds to the color singlet contribution. 
We have compared our calculation with the recent PHENIX data (run3). 
The daggers are recent PHENIX data taken from \cite{phenix}. 
It can be seen from the figure that the color singlet contriubtion is more than an order of magnitude 
smaller than the octet contribution. On the other hand, the
color octet mechanism with only parton fusion processes reproduces the recent PHENIX data very well. Thus for $p_T \leq 5 GeV $ 
we do not find it necessary to include contributions from 
the parton fragmentation processes.  
We have not compared our calculation with the data below $p_T$ equals to 2 
GeV as pQCD calculations are not valid at such low transverse momenta. 
In the near future PHENIX will have sensitivity 
up to  $p_T = 10 GeV$. This new data will allow us to decide whether 
fragmentation processes contributions within color octet mechanism are 
important at higher $p_T$ and whether their inclusion will allow us to 
understand  the data.

In Fig. 2 we present the rapidity distribution of the $J/ \psi$ production 
cross section in pp collisions at RHIC at $\sqrt s$ = 200 GeV. We have taken the
charm quark mass equal to 1.45 GeV in our calculation. 
The solid line is the results obtained by using the GRV98 PDF's with lowest values of the
non-perturbative matrix elements.
The upper dashed line corresponds to the results obtained by using the GRV98 PDF's with highest 
values of the
non-perturbative matrix elements. 
The upper dot-dashed line corresponds to the results obtained by using the MRST99 PDF's with highest values of the
non-perturbative matrix elements. 
The lower dot-dashed line corresponds to the results obtained by using the MRST99 PDF's with lowest values of the
non-perturbative matrix elements. 
The PHENIX experimental data \cite{phenix} are shown in the figure 
as daggers. 
The rapidity range covered by PHENIX at RHIC is from $-3 < y < 3$. 
The color singlet contribution is also shown as the lower dashed line (here we only show the results using 
the  GRV98 PDF's)  which is very small when compared with the 
color octet contribution. It can be seen that just including the color octet contribution 
explains the run3 PHENIX data.

Finally, we present the total cross section  for  $J/\psi$  production in pp 
collisions at $\sqrt s$ = 200 GeV at RHIC and compare it with the run3 
PHENIX data. 
In the color octet case we find $B \sigma_{J/\psi}$= 159 $\pm$ 6 (nb)
which is in very good agreement with the
PHENIX collaoration experimental value which is reported as 159 (nb) \cite{phenix}. The $\pm$ 6
(nb) is due to the errors in the non-perturbative matrix elements as given in \cite{cho2}.
In the above total cross section computation we have used charm quark mass 
equal to 1.45 GeV and the MRST99 parton distribution function inside the proton. 
The agreement of color octet mechanism predictions with 
the 
PHENIX experimental data is remarkable in all aspects, {\it i.e.} 
in $p_T$ distribution, 
in rapidity distribution and in total cross section measurements at $\sqrt s$ = 200 GeV pp collisions. 

Before summarizing, we briefly discuss the uncertainties due to PDF's, non-perturbative matrix elements
etc. As can be seen from Fig. 1 and Fig. 2 that both GRV98 and
MRST99 PDF's describe the data very well. 
We have also checked that
the recent CTEQ6M PDF's also explain the data very well. The uncertainty in the results due to the
errors in the non-perturbative matrix elements are given in Fig. 1 and 2 and also in the total cross section.

In summary, in this letter, we have studied  $J/\psi$  production in pp collisions at RHIC at $\sqrt s$ = 
200 GeV within the 
PHENIX detector acceptance range by using the color singlet and the
color octet mechanism within pQCD and NRQCD and have compared them with the 
recent 
PHENIX experimental data.We have demonstrated  that the color octet mechanism using only parton fusion processes is able to
 reproduce the RHIC data on the $p_T$ distribution, 
rapidity distribution and total cross section of  $J/\psi$  
production. This is done without any normalization factor modifications and using matrix elements
extracted from the Tevatron and other experiments. 
 The color singlet mechanism gives negligible contributions when compared to the octet contributions at this energy.  Our results confirm the applicability of the color octet mechanism found at the Tevatron. 

 It is important to identify the correct 
mechanism for  $J/\psi$  production in pp collisions 
at RHIC if we want to 
study  $J/\psi$  suppression as a signature of quark-gluon plasma 
detection in AuAu collisons 
at $\sqrt s$ = 200 GeV \cite{nayak}.  It is also important if we hope 
to extract  the polarized gluon distribution function from the data on polarized 
pp collisions at RHIC.  The latter will be important for isolating interference terms 
which can be used to explore various models which predict non standard model TeV physics.

\acknowledgements {\bf Acknowledgements:}
We thank Rajiv Gavai, Sourendu Gupta, Pat McGaughey, Jack Smith and George Sterman
for useful discussions. This work was supported in part by the National Science
Foundation grant PHY-0098527 and Depertment of Energy, under contract W-7405-ENG-36.

\begin{figure}
\centering
\includegraphics[width=3in]{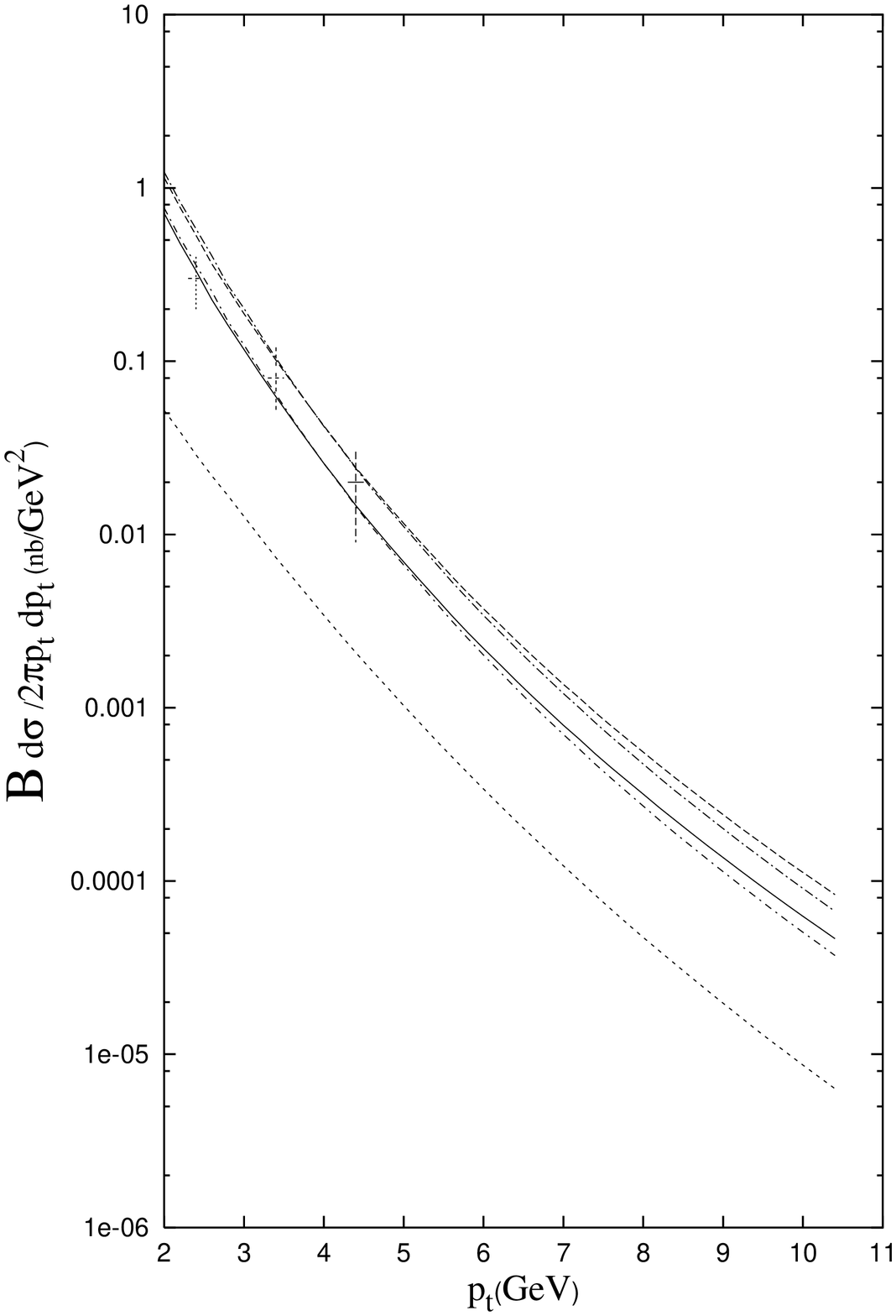}
\caption{
$p_T$ distribution of $J/\psi$ production cross section 
in pp collisions at $\sqrt s$ = 200 GeV 
at RHIC in the PHENIX detector acceptance range. The solid and upper dashed lines and upper/lower
dot-dashed lines correspond to color octet mechanism predictions with GRV98 and MRST99 PDF's. The
lower dashed line is the color singlet contribution. The daggers 
are run3 PHENIX 
experimental data. 
} 
\label{fig1}
\end{figure}

\begin{figure}
\centering
\includegraphics[width=3in]{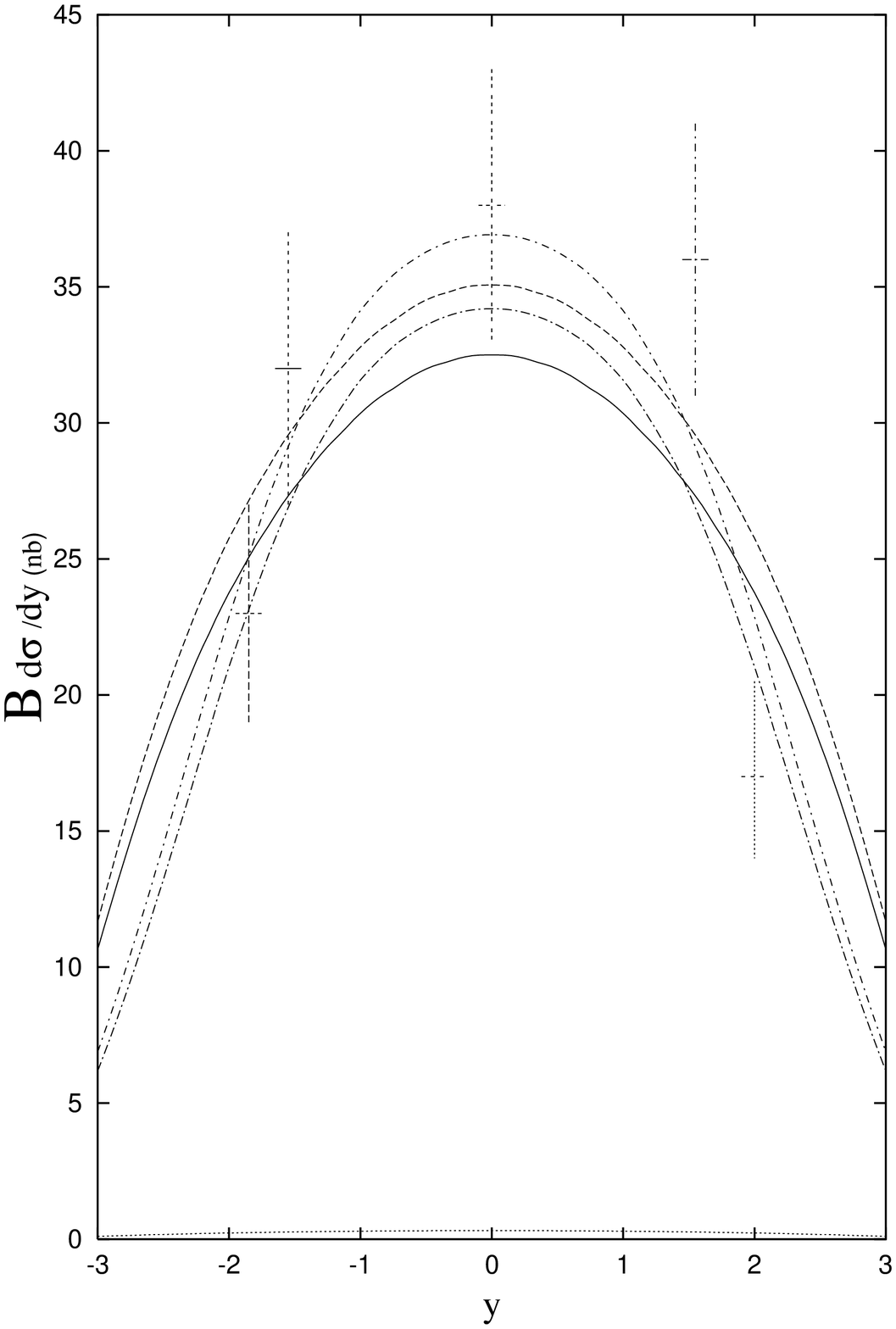}
\caption{
Rapidity distribution of $J/\psi$
production cross section 
in pp collisions at $\sqrt s$ = 200 GeV at RHIC in the PHENIX 
detector 
acceptance range. The solid and upper dashed lines and upper/lower
dot-dashed lines correspond to color octet mechanism predictions with GRV98 and MRST99 PDF's. The
lower dashed line is the color singlet contribution. The daggers are run3 PHENIX experimental data. 
} 
\label{fig2}
\end{figure}

\end{document}